\documentclass[sigconf,natbib=true]{acmart}
\usepackage{graphicx}
\usepackage{multirow}
\usepackage{subcaption}
\usepackage{bm}

\AtBeginDocument{%
  \providecommand\BibTeX{{%
    \normalfont B\kern-0.5em{\scshape i\kern-0.25em b}\kern-0.8em\TeX}}}

\copyrightyear{2024}
\acmYear{2024}
\setcopyright{acmlicensed}\acmConference[KDD '24]{Proceedings of the 30th ACM SIGKDD Conference on Knowledge Discovery and Data Mining}{August 25--29, 2024}{Barcelona, Spain}
\acmBooktitle{Proceedings of the 30th ACM SIGKDD Conference on Knowledge Discovery and Data Mining (KDD '24), August 25--29, 2024, Barcelona, Spain}
\acmDOI{10.1145/3637528.3671973}
\acmISBN{979-8-4007-0490-1/24/08}

\begin{document}

\title{When Box Meets Graph Neural Network in Tag-aware Recommendation}


\author{Fake Lin}
\email{fklin@mail.ustc.edu.cn}
\orcid{0009-0003-1402-2358}
\affiliation{%
  \institution{University of Science and Technology of China}
  \city{Hefei}
  \country{China}
}

\author{Ziwei Zhao}
\email{zzw22222@mail.ustc.edu.cn}
\orcid{0009-0007-0814-7313}
\affiliation{%
  \institution{University of Science and Technology of China}
  \city{Hefei}
  \country{China}
}

\author{Xi Zhu}
\email{xizhu@mail.ustc.edu.cn}
\orcid{0000-0003-3621-8493}
\affiliation{%
  \institution{University of Science and Technology of China}
  \city{Hefei}
  \country{China}
}
\author{Da Zhang}
\email{zhangda17@mail.ustc.edu.cn}
\orcid{0009-0004-4247-2468}
\affiliation{%
  \institution{University of Science and Technology of China}
  \city{Hefei}
  \country{China}
}

\author{Shitian Shen}
\email{shitians@gmail.com}
\orcid{0000-0001-6859-5758}
\affiliation{%
  \institution{Alibaba Group}
  \city{Hangzhou}
  \country{China}
}

\author{Xueying Li}
\email{xiaoming.lxy@alibaba-inc.com}
\orcid{0000-0002-3699-0697}
\affiliation{%
  \institution{Alibaba Group}
  \city{Hangzhou}
  \country{China}
}

\author{Tong Xu}
\authornote{Corresponding Author.}
\email{tongxu@ustc.edu.cn}
\orcid{0000-0003-4246-5386}
\affiliation{%
  \institution{University of Science and Technology of China}
  \city{Hefei}
  \country{China}
}
\author{Suojuan Zhang}
\email{suojuanzhang@aeu.edu.cn}
\orcid{0009-0003-2193-2288}
\affiliation{%
  \institution{Army Engineering University of PLA}
  \city{Nanjing}
  \country{China}
}
\author{Enhong Chen}
\authornotemark[1]
\email{cheneh@ustc.edu.cn}
\orcid{0000-0002-4835-4102}
\affiliation{%
  \institution{University of Science and Technology of China}
  \city{Hefei}
  \country{China}
}

\renewcommand{\shortauthors}{Fake Lin and Ziwei Zhao, et al.}

\begin{abstract}
Last year has witnessed the re-flourishment of tag-aware recommender systems supported by the LLM-enriched tags. Unfortunately, though large efforts have been made, current solutions may fail to describe the diversity and uncertainty inherent in user preferences with only tag-driven profiles. Recently, with the development of geometry-based techniques, e.g., box embedding, diversity of user preferences now could be fully modeled as the range within a box in high dimension space. However, defect still exists as these approaches are incapable of capturing high-order neighbor signals, i.e., semantic-rich multi-hop relations within the user-tag-item tripartite graph, which severely limits the effectiveness of user modeling. To deal with this challenge, in this paper, we propose a novel algorithm, called BoxGNN, to perform the message aggregation via combination of logical operations, thereby incorporating high-order signals. Specifically, we first embed users, items, and tags as hyper-boxes rather than simple points in the representation space, and define two logical operations to facilitate the subsequent process. Next, we perform the message aggregation mechanism via the combination of logical operations, to obtain the corresponding high-order box representations. Finally, we adopt a volume-based learning objective with Gumbel smoothing techniques to refine the representation of boxes. Extensive experiments on two publicly available datasets and one LLM-enhanced e-commerce dataset have validated the superiority of BoxGNN compared with various state-of-the-art baselines. The code is released online \footnote{https://github.com/critical88/BoxGNN}.
\end{abstract}

\ccsdesc[500]{Information systems~Recommender systems}

\keywords{Tag-aware Recommendation, Box Embedding, Graph Neural Networks}



\maketitle

\section{INTRODUCTION}
\begin{figure}[t]
\centering
\includegraphics[scale=0.14 ]{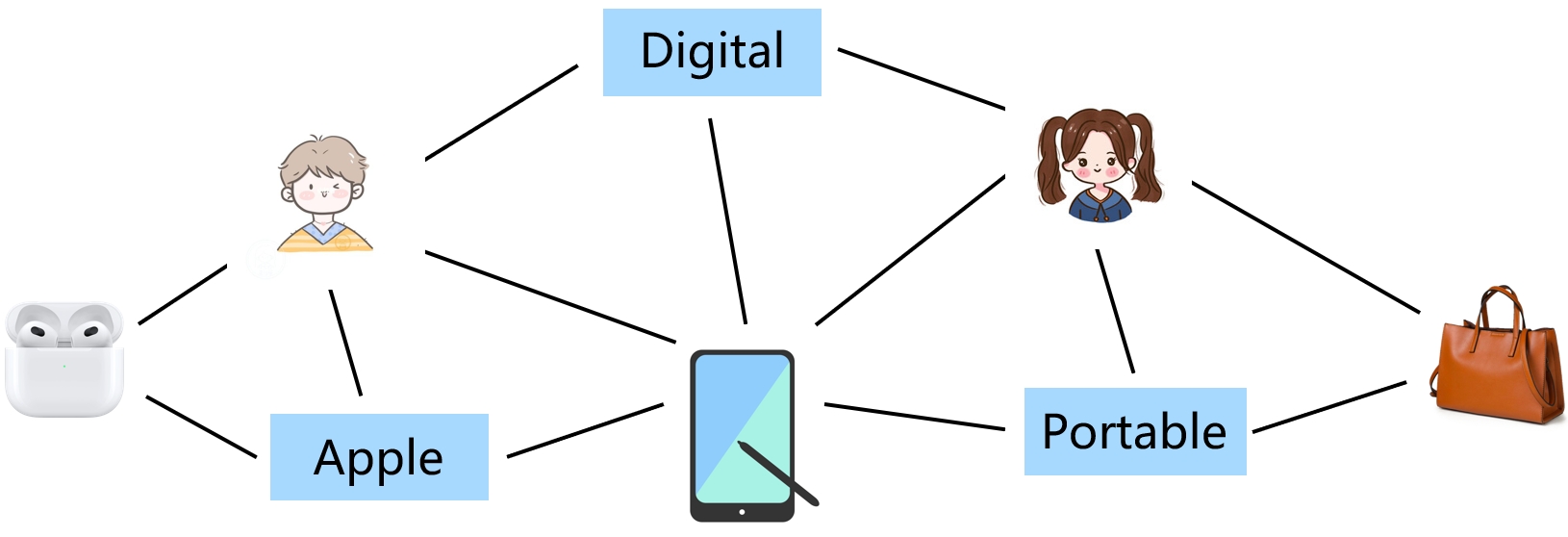} 
\caption{An example of collaborative tag graph that users, items and tags are interconnected with each other. It illustrates that users may buy items due to the diverse interests.}
\label{fig0}
\end{figure}

Tag-aware Recommender System (TRS) has long been treated as a crucial foundation to support the intelligent e-commerce platforms, especially with the support of semantic-rich tags generated by large language models (LLMs) \cite{wu2023survey}. Along this line, it is necessary to capture the tag semantics for building user profiles. Traditionally, prior arts could be roughly divided into two categories. The first line of literatures could be \emph{feature-based}, which mainly focus on encoding tags as multi-hot vectors that can be easily processed by following-up applications~\cite{DBLP:journals/ijon/ZuoZGJ16,DBLP:journals/ijon/HuangWHYC20,DBLP:journals/ijon/ChenDXLWW21}. Unfortunately, they may suffer from sparsity issue, making it difficult to depict preferences of users who are inactive in the platform. More importantly, they usually ignore the high-order signals, i.e., semantic-rich multi-hop relations within the user-tag-item tripartite graph, which severely limits the effectiveness of user profiling. To that end, some other researches attempt to exploit the rich semantics within high-order signals via graph neural network (GNN) techniques~\cite{DBLP:conf/cikm/ChenGTXDHW20,DBLP:conf/ijcnn/HuangHC21, DBLP:journals/corr/abs-2208-03454,chen2024macro}, which perform the message aggregation mechanism to capture high-order interactions from multi-hop neighbors, thereby improving recommendation quality. However, these approaches usually represent users, items, and tags as fixed points within shared vector space. In this case, the diversity and uncertainty inherent in user preferences, i.e., the same user may exhibit varied preferences in different cases, could not be well represented. For instance, From Fig. \ref{fig0}, we can observe that the reasons behind their purchases of the "iPhone" can also differ: a boy might buy it for its "Apple" brand, whereas a girl might choose it for its portability. In summary, more comprehensive solution for TRS task is still urgently required.
Recently, with the development of geometric embeddings, e.g., \textbf{box embedding} in high-dimensional space, it is possible to describe the diversity and uncertainty of user preferences~\cite{DBLP:conf/sigir/LiangZDXLY23} via the intersection of high-dimensional boxes which encodes users/items, respectively. Moreover, the box embedding mechanism could well fit the hierarchical structure of tagging system~\cite{DBLP:conf/acl/OnoeBMD20}, like “football” as a subset of “sport”, which further enrich the semantic of tag representations. Nevertheless, these methods only examine the direct interactions (e.g., purchase record) among items and users, neglecting the importance of collaborative signals from higher-order neighbors for revealing user preferences~\cite{DBLP:conf/sigir/0001DWLZ020,zheng2023interaction,DBLP:journals/csur/WuSZXC23,LAGCN,DBLP:conf/www/Chen0SS0Z024}. 

\begin{figure}[t]
\centering
\includegraphics[scale=0.7 ]{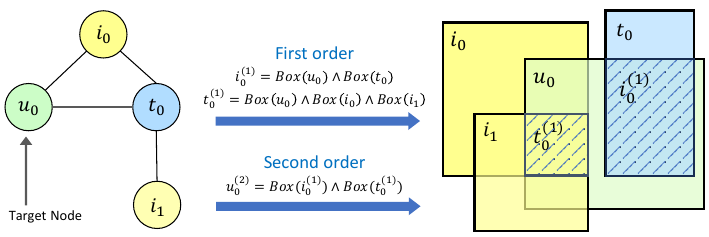} 
\caption{Toy example of message aggregation via intersection operation, where $Box(\cdot)$ represents the box embedding of nodes, $u$, $i$ and $t$ denotes user, item and tag, respectively. Notably, we fail to obtain the 2nd-order user box embedding, as  $t_0^{(1)}$ and $i_0^{(1)}$ are disjoint.}
\label{fig1}
\end{figure}

In this paper, we aim to capture high-order collaborative signals for solving the TRS task, while preserving the powerful representational capability of the box embedding mechanism. In detail, combination of logical operations on embedded boxes will be utilized to simulate the message aggregation process of graphs. Along this line, there are two challenges urged to be addressed:
\begin{enumerate}
    \vspace{-1mm}
    \item \textbf{How to aggregate the box embeddings}? The learned box embeddings tend to be anisotropic~\cite{DBLP:conf/sigir/LiangZDXLY23}, i.e., the length ranges of different box dimensions vary greatly, which could easily lead to an empty set when discovering the intersection of two embedded boxes. Therefore, it is intractable to use interaction operation. To make matters even worse, stacking multiple layers will further exacerbate this issue. For instance, as depicted in Fig.\ref{fig1}, by solely utilizing intersections for message aggregation, we can easily obtain the first-order aggregated nodes $i_0^{(1)}$ and ${t_0^{(1)}}$ \cite{DBLP:conf/sigir/LiangZDXLY23}. However, it is impractical to obtain the 2nd-order user box $u_0^{(2)}$, because no overlapping parts between $i_0^{(1)}$ and ${t_0^{(1)}}$ can be found in the vector space. In summary, we need to devise an innovative strategy to aggregate embedded boxes.
    \item \textbf{How to measure the matching score between user and item?} To the best of our knowledge, conventional approaches \cite{DBLP:journals/corr/abs-1205-2618,DBLP:conf/sigir/0001DWLZ020,DBLP:conf/cikm/MaoZWDDXH21} typically use dot product or cosine similarity to calculate the matching degree between users and items. However, these methods could be hardly transferred to box embedding scenario, as they could only reflect the information of box center, but not the wide range of the whole box. Recent box-based studies, e.g., ~\cite{DBLP:conf/sigir/LiangZDXLY23,DBLP:conf/sigir/ChenYLNWW22} typically employ a distance formula to compute the scores between users and items. Nonetheless, it is unwise that the distance between two boxes serves as their similarity, as it would overlook rich information in the overlapping areas to represent the preference uncertainty. Consequently, it seems feasible to utilize the volume of the intersection areas as the matching score. In this line, we may face the gradient vanish issue if there is no overlap between the boxes. In summary, we are expected to formulate a smoothed volume-based methodology to consistently deliver gradient signals.
\end{enumerate}

To tackle the challenges, we propose a novel tag-aware recommendation algorithm, named \textbf{BoxGNN}, aiming to harness the high-order collaborative signals based on the embedded boxes. Specifically, we first transform the users, items and tags into separate box embedding, and then implement two operations, i.e. intersection and union, to facilitate the following process. Afterwards, upon thoroughly assessing the role of each type (a.k.a. user, item and tag), we tailor three combinations of operations instead of relying solely on the intersection operation. In this way, we can seamlessly aggregate messages of graph neighbors and obtain the high-order box embeddings. Finally, to avoid gradient vanishing issue, we employ a Gumbel technique to smooth the gradient signal and ultimately make the proposed model trainable. Technical contribution of this paper could be summarized as follows:
\begin{itemize}
\item We propose an novel solution for the TRS task, which adapt graph neural network into the box embedding mechanism to derive high-order signals, while capturing uncertainty of user preference simultaneously.
\item We apply three strategies to perform message aggregation based on the type of box (i.e. user, item and tag). Furthermore, gumbel smoothing technique is utilized to ensure the differentiability during training process.
\item We conduct extensive experiments on two public benchmarking datasets and one LLM-enhanced dataset to justify the effectiveness of our proposed BoxGNN. 
\end{itemize}



\section{RELATED WORK}

\subsection{Tag-Aware Recommendation}
As its powerful capability to express user interests, large efforts have been devoted to tag-aware recommendation. Initially, researchers extend the traditional collaborative approaches to incorporate information from tags \cite{DBLP:conf/sac/Tso-SutterMS08,DBLP:conf/gfkl/MarinhoS07,DBLP:conf/ijcai/XuLCMM17,DBLP:conf/bcd/Maleki-ShojaT19}. To alleviate the redundancy of the tag information, cluster-based schemes \cite{DBLP:conf/recsys/ShepitsenGMB08} were proposed to merge the uninformative tags. With the prevalence of deep learning, \cite{DBLP:journals/ijon/ZuoZGJ16} tries to learn tag-enhanced user representation via deep autoencoders and employ collaborative filter to arrive at recommendations. Moreover, AIRec\cite{DBLP:journals/ijon/ChenDXLWW21} developed a hierarchical attention network to capture multifaceted user representations and introduced an intersection module designed to derive conjunctive features from the overlap between user and item tags.

Recently, GNN-based approaches have shown its superiority in the tag-aware recommendation systems \cite{DBLP:conf/cikm/ChenGTXDHW20,DBLP:journals/corr/abs-2208-03454}. These models first build an unified graph to express the connectivity among users, items and tags. Then they aggregate messages from neighbors to enrich the ego embeddings, which incorporates the tag information into collaborative filters  to facilitate the final recommendation. Specifically, TGCN \cite{DBLP:conf/cikm/ChenGTXDHW20} employs a novel message propagation approach to model tag information, thereby recognizing user interests at multiple granularities. LFGCF \cite{DBLP:journals/corr/abs-2208-03454} borrow the idea from the LightGCN \cite{DBLP:conf/sigir/0001DWLZ020} to learn the high-order representations of users, items and tags, which boosts the recommendation performance.
Although their effectiveness, the uncertainty of user interests and the hierarchical nature of tags remain unexplored, which results in the suboptimal performance.

\subsection{Geometric Embedding}
Geometric embedding technique have garnered widespread attention due to the ability to preserve the intrinsic geometric structure of data. These techniques map high-dimensional information into a lower-dimensional space while retaining the significant relationships and topological features of the original data \cite{DBLP:conf/iclr/RenHL20,DBLP:conf/sigir/LiangZDXLY23,DBLP:conf/acl/DasguptaBMAP0M22,DBLP:conf/nips/DasguptaBZV0M20}. At the beginning, some researchers simply used geometric embedding to represent the implicit partial order relations in data \cite{DBLP:journals/corr/VendrovKFU15}. Then, this line of literature exploits its strong capability to model complex logical structural information in knowledge graph \cite{DBLP:conf/iclr/RenHL20}. Next, more sophisticated models have been developed to adapt to a variety of scenarios and datasets \cite{DBLP:conf/nips/ZhangWCJW21,DBLP:conf/akbc/PatelDB0VM20,DBLP:conf/nips/AbboudCLS20}.

Recently, researchers extend this modeling techniques to the recommendation systems \cite{DBLP:conf/sigir/LiangZDXLY23,DBLP:conf/sigir/ChenYLNWW22}. Specifically, CubeRec \cite{DBLP:conf/sigir/ChenYLNWW22} considers the groups to be the hypercubes to resolve the long-standing issue of data sparsity and employs the self-supervision to learn the expressive representations of groups.
CBox4cr \cite{DBLP:conf/sigir/LiangZDXLY23} tries to capture the user interest by the intersection of the item sequence that user interacted with. Although their achievement, they still ignore the high-order signals in recommendation system, which is proven effective in various recommendation scenarios \cite{DBLP:conf/sigir/0001DWLZ020,zheng2021drug,DBLP:conf/sigir/Wang0WFC19,ALDI,DBLP:conf/dasfaa/ZhuLZXLYLC23,DBLP:conf/sigir/ZhaoZXLYLYC23}.

\section{PRELIMINARIES}
In this section, we first introduce the formal formulation of tag-aware recommendation task, and then proceed to the definition of collaborative tag graph (CTG) to support the user profiling.
\subsection{Task Formulation}
Social tagging system encourages user to assign a user-defined tag to the items of interest. Therefore, tags implicitly serve as the user interest in the items. Following prior works, suppose there exists user set $\mathcal{U}$, item set $\mathcal{I}$ and tag set $\mathcal{T}$, where their size are $|\mathcal{U}|=N_u$, $|\mathcal{I}|=N_i$ and $|\mathcal{T}|=N_t$, respectively.
Then a tagging assignment can be formulated as a triplet, i.e. $a=(u,t,i)$, which denotes that a user $u$ annotates a tag $t$ to the item $i$. Similar to \cite{DBLP:conf/cikm/ChenGTXDHW20}, the folksonomy is a tuple $\mathcal{F}=(\mathcal{U}, \mathcal{I}, \mathcal{T}, \mathcal{A})$, where $\mathcal{A} \in \mathcal{U} \times \mathcal{I} \times \mathcal{T}$ is the assignments in typical tag-aware recommendation.

Following \cite{DBLP:conf/cikm/ChenGTXDHW20}, we consider tags as a complement to the information of users and items. Here we target at predicting the interaction between users and items, which is represented by $Y \in \{0,1\}^{N_u \times N_i} $, where each entry $y_{ui} = 1$ indicates that the given user $u \in \mathcal{U}$ has interacted with item $i \in \mathcal{I}$, otherwise $y_{ui} =0$. Here comes the formal definition of tag-aware recommendation task:

\begin{definition}
    \textbf{Tag-aware Recommendation System (TRS)}. Given the user set $\mathcal{U}$, the item set $\mathcal{I}$, and their observed interactions (annotations) $Y$, with the complementary information from $\mathcal{T}$, TRS aims to learn a model that is capable of predicting the top-k item list that meets the interests for each user $u \in \mathcal{U}$. 
\end{definition}
\subsection{Collaborative Tag Graph}
To ease the understanding in overall process of message aggregation, we build a unified graph which includes three type of nodes, i.e., user, item and tags. Here is the definition: 
\begin{definition}
\textbf{Collaborative Tag Graph.} A collaborative tag graph is an undirected graph $\mathcal{G}=(\mathcal{V}, \mathcal{E})$, where $\mathcal{V} \in \mathcal{U} \cup \mathcal{I} \cup \mathcal{T}$ denotes the node set that includes three types of nodes and $\mathcal{E}$ represents the edges. It is worth noting that an assignment $(u, t, i)$ will be splitted into three edges, that is $(u,i)$,$(u,t)$ and $(i,t)$. For simplicity, we define the relation set as $\mathcal{R} \in \{r_0,r_1,r_2\}$, where $r_0$, $r_1$, $r_2$ represents the relation between $(u,i)$, $(u,t)$ and $(i,t)$, respectively. Finally, the collaborative tag graph can be formally described as:
\begin{equation}
    \begin{aligned}
        \mathcal{G}&=\{(v,r,v')| v,v' \in \mathcal{V}, r \in \mathcal{R}\},
    \end{aligned}
\end{equation}
\end{definition}

\section{METHODOLOGY}
\begin{figure*}[htbp]
\centering
\includegraphics[scale=0.95]{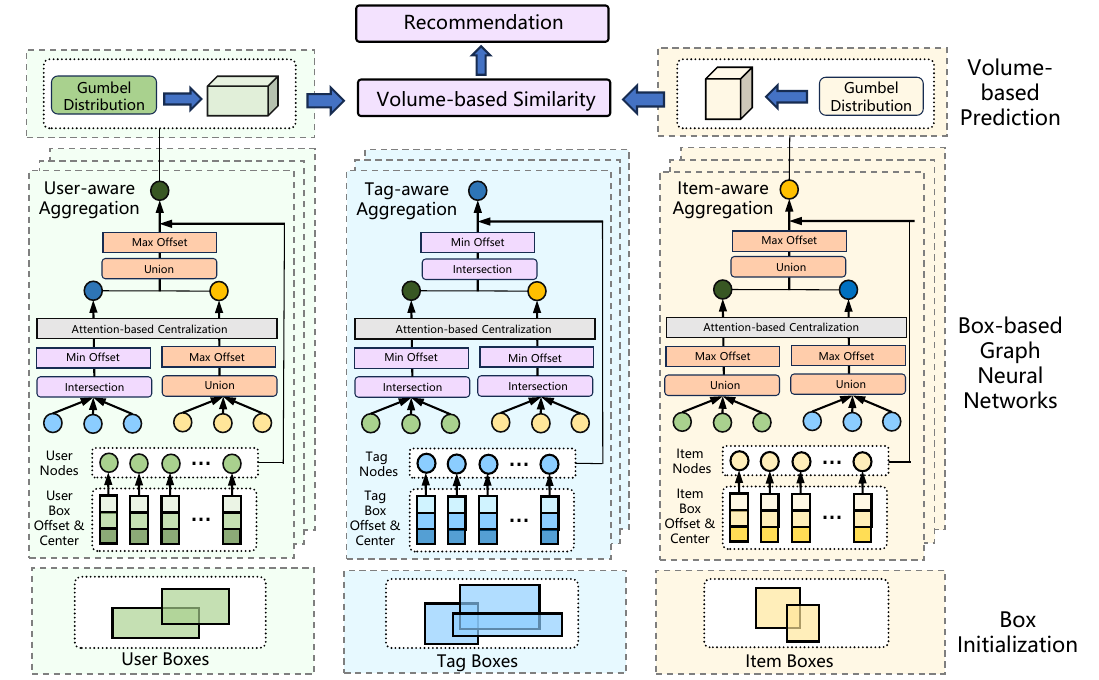} 
\caption{The overall framework of the proposed BoxGNN. Note the Intersection and Union are logical operations including Min/Max Offset and Centralization, where Min Offset is to obtain the minimum offset among boxes in an element-wise manner. } 
\label{framework}
\end{figure*}

In this section, we elaborate on the overall process of BoxGNN, which consists of three steps: (a) We firstly introduce the formulation of box embedding, and transform all nodes into box representation. (b) Subsequently, we perform the message aggregation in the context of box language to obtain the high-order representations. (c) Eventually, we utilize the Gumbel-based volume of the intersection between arbitrary two boxes to guide the whole learning process. The framework of BoxGNN is depicted in Fig \ref{framework}. 

\subsection{Box Initialization}
In this part, we introduce the approach of representing nodes as boxes in a multi-dimensional space and formulate the logical operations that apply to these boxes in subsequent sections.


\subsubsection{Box Embedding} 

Different from earlier models \cite{DBLP:conf/sigir/LiangZDXLY23} that only portrayed items as boxes, we extend this modeling to all nodes within the CTG as boxes. This change provides us with a unified language for logical operations across the CTG and aligns with our objectives of capturing user interest uncertainty. Formally, a box is defined as $\boldsymbol{p}\equiv (Cen(\boldsymbol{p}), Off(\boldsymbol{p})) \in \mathbb{R}^{2d}$, then:
\begin{equation}
    \begin{aligned}
        Box_{\boldsymbol{p}}\equiv  \{\boldsymbol{v}\in \mathbb{R}^d: Cen(\boldsymbol{p}) - Off(\boldsymbol{p})\preceq \boldsymbol{v} \preceq Cen(\boldsymbol{p}) + Off(\boldsymbol{p})\}
    \end{aligned}
\end{equation}

\noindent where $\preceq$ denotes the element-wise inequality, $Cen(\boldsymbol{p})$ and $Off(\boldsymbol{p})$ are the center and the offset of the box $\boldsymbol{p}$, respectively. For convenience, we denote the box representation of a user, item, tag and node as $\boldsymbol{u}$, $\boldsymbol{i}$, $\boldsymbol{t}$ and $\bm{v}$, respectively.

\subsubsection{Logical Operations}

After initializing the box representations for all nodes, it is impractical to apply traditional operators such as addition and multiplication. Therefore, we need to define a set of logical operations for boxes. Without loss of generality, we have two logical operations: intersection and union. Here comes their specific definition:

\textbf{Intersection.} Given a node set $\{v_1, v_2, \dots, v_n\}\in \mathcal{V}$, intersection operation obtains the $\boldsymbol{v}_{inter}=\bigcap_{k=1}^n \boldsymbol{v}_k$, where $\boldsymbol{v}_k$ is the corresponding box representation and $\boldsymbol{v}_{inter}$ denotes the intersected box, which is defined as $\boldsymbol{v}_{inter}=(Cen(\boldsymbol{v}_{inter}), Off(\boldsymbol{v}_{inter}))$. The $Cen(\boldsymbol{v}_{inter})$ and $Off(\boldsymbol{v}_{inter})$ are derived as follows:
\begin{equation}
    \begin{aligned}
        Cen(\boldsymbol{v}_{inter})=\sum_{i = 1}^n {a_i \odot Cen(\boldsymbol{v}_i)},a_i=\frac{exp(MLP(Cen(\boldsymbol{v}_i)))}{\sum_{j=1}^n{exp(MLP(Cen(\boldsymbol{v}_j)))}},
    \end{aligned}
\end{equation}
\begin{equation}
    \begin{aligned}
    \label{equation_2}
        Off(\boldsymbol{v}_{inter})=Min(\{Off(\boldsymbol{v}_1),\dots,Off(\boldsymbol{v}_n)\})
    \end{aligned}
\end{equation}
where $\odot$ is the element-wise product, $MLP$ is the Multi-Layer Perceptron, $Min$ and $exp$ are used in element-wise manner. \\

\textbf{Union.} Similar to intersection, given the node set $\{v_1, \dots, v_n\} \in \mathcal{V}$, we can still generate a new box representation $\boldsymbol{v}_{uni}=\bigcup_{k=1}^n \boldsymbol{v}_k$, where $\boldsymbol{v}_{uni}\equiv(Cen(\boldsymbol{v}_{uni}), Off(\boldsymbol{v}_{uni}))$, which is generated by performing attentional sum over the box centers and expanding the box offset:
\begin{equation}
    \begin{aligned}
        Cen(\boldsymbol{v}_{uni})=\sum_{i=1}^n {a_i \odot Cen(\boldsymbol{v}_i)},a_i=\frac{exp(MLP(Cen(\boldsymbol{v}_i)))}{\sum_{j=1}^n{exp(MLP(Cen(\boldsymbol{v}_j)))}},
    \end{aligned}
\end{equation}
\begin{equation}
    \begin{aligned}
        Off(\boldsymbol{v}_{uni})=Max(\{Off(\boldsymbol{v}_1),\dots,Off(\boldsymbol{v}_n)\})
    \end{aligned}
\end{equation}
Upon acquiring these two operators, it becomes imperative to delve into their roles within tag-aware recommendation. The intersection operator preserves the overlapping areas across multiple boxes, which embodies the shared attributes and characteristics. For instance, given the item set in the user interaction histories, the intersection of them reveals their shared traits, which indicates the implicit interests of the user. Conversely, the union operation retains all the information from the boxes, allowing the aggregated box to effectively incorporate information from boxes. For example, considering an item with abundant tags, performing union operation on these tags yields a box with comprehensive semantic information from the tags.

\subsection{Box-based Graph Neural Network}
\label{section:gnn}
In this subsection, we apply the language of boxes to interpret the message-passing mechanism to capture high-order signals. In traditional GNN frameworks \cite{DBLP:conf/sigir/0001DWLZ020,DBLP:journals/csur/WuSZXC23}, we typically use mean or weighted sum to aggregate neighboring information. However, in the context of boxes, only intersection and union can be utilized to aggregate neighboring information. 
This constraint requires us to rethink the aggregation process in GNNs, as these operations are fundamentally different from arithmetic aggregations and can capture the complex relationships and structures that may exist among the nodes in CTG. Specifically, different nodes may require distinct strategies when encountering various types of neighbors. Therefore, we conduct an in-depth analysis of each type of node and provide the corresponding aggregation formulas.

\textbf{User-aware Aggregation.}
The key to the recommendation system lies in modeling user interests. Following previous studies, we aggregate the neighbors of the user to obtain the higher representations. However, there exists two type of neighbors, i.e., tags and items, which may exert different influences on the user. Along this line, we have divided the propagation process into two parts for separate study.

For tag side, users are assigned a large number of tags, manifested as edges between users and tags in CTG. These tags can be considered as explicit interests of the users, with multiple interests overlapping, thus forming a comprehensive representation of user interests. In the context of box language, the user box representation can be aggregated by the box representations of multiple tags, i.e., through a union operation: 

\begin{equation}
    \begin{aligned}
        \boldsymbol{u}^{(l+1)}_t = \bigcup_{k=1}^ {|\mathcal{N}_u^t|}\boldsymbol{t}_k^{(l)}, \\
    \end{aligned}
\end{equation}
where $\boldsymbol{u}^{(l+1)}_t$ is the $(l+1)$-th user box representation for aggregating tag neighbors and $\mathcal{N}_u^t$ is the tag neighbor set of the user $u$. 

Compared with the tag, item neighbors may tell the different story to the user. Notably, item neighbors are the history interactions of the user, which indicates that these purchased items provide insight into user preferences. Therefore, it is rational to extract user profiles from its purchased items, which can be interpreted as performing an intersection operation on these item boxes. Along this line, the conjunctive box expresses the shared traits in items, which indicates the user implicit preference:
\begin{equation}
    \begin{aligned}
        \boldsymbol{u}^{(l+1)}_i = \bigcap_{k=1}^ {|\mathcal{N}_u^i|}\boldsymbol{i}_k^{(l)}, \\
    \end{aligned}
\end{equation}
where $\boldsymbol{u}^{(l+1)}_i$ is the user box representation aggregated by item neighbors and $\mathcal{N}_u^i$ is the item neighbor set of the user $u$. Finally, by union the interests from two aspects, we obtain the final user box: 

\begin{equation}
    \begin{aligned}
        \label{eq:cen}
        \boldsymbol{u}^{(l+1)} &= \boldsymbol{u}_t^{(l+1)} \cup \boldsymbol{u}_i^{(l+1)}, \\
        Cen(\boldsymbol{u}^{(l+1)}) & = \sum_{k=1}^{|\mathcal{N}_u|} {a_k \odot Cen(\boldsymbol{v}_k^{(l)})}, \\
        a_k & = \frac{exp(MLP(Cen(\boldsymbol{v}_k^{(l)})))}{\sum_{j=1}^{|\mathcal{N}_u|}{exp(MLP(Cen(\boldsymbol{v}_j^{(l)})))}}, \\
        Off(\boldsymbol{u}^{(l+1)})&= Max(Min(\{Off(\boldsymbol{v}_i^{(l)})\}), Max(\{Off(\boldsymbol{v}_t^{(l)})\}))
    \end{aligned}
\end{equation}
where $\mathcal{N}_u$ is the all neighbors of the user $u$, $\boldsymbol{v}_k$ denotes the tag or item box representations. In the following formulation, we omit the derivation of $Cen(\cdot)$, as they is the same as depicted in Eq. \ref{eq:cen}.

\textbf{Tag-aware Aggregation.}  Tags serve as an additional information to enrich the user and item embeddings. In TRS scenario, tag plays crucial role in connecting two users with similar interests or items with shared characteristics. Therefore, the neighbors for the tag are users and items, and we study the impact of these two types on the tag separately.

For user, a tag can be assigned to multiple users, representing their common interests and hobbies. However, due to the diversity of user interests, the tags can represent a small part of user interests, which is hard to be located. Therefore, we apply intersection to extract the shared information from the intricate representations of the neighboring users, which is referred as the high-order representation of the tag. This can be formalized as follows:
\begin{equation}
    \begin{aligned}
        \boldsymbol{t}^{(l+1)}_u = \bigcap_{k=1}^ {|\mathcal{N}_t^u|}\boldsymbol{u}_k^{(l)}, \\
    \end{aligned}
\end{equation}
where $\boldsymbol{t}^{(l+1)}_u$ is the $(l+1)$-th tag box representation for aggregating user neighbors and $\mathcal{N}_t^u$ is the user neighbor set of the tag $t$. 

Similar to user, a tag is assigned to multiple items which implies their attributes or features. In this case, the key to obtain the high-order representation of tags still lies in the distillation of the shared features in the item neighbors. Likewise, we apply intersection to aggregate the information from item neighbors:
\begin{equation}
    \begin{aligned}
        \boldsymbol{t}^{(l+1)}_i = \bigcap_{k=1}^ {|\mathcal{N}_t^i|}\boldsymbol{i}_k^{(l)}, \\
    \end{aligned}
\end{equation}
where $\boldsymbol{t}^{(l+1)}_i$ is the tag box representation aggregated by item neighbors and $\mathcal{N}_t^i$ is the item neighbor set of tag $t$. Finally, by combining these two parts, we can obtain the final high-order box representation of the tag:
\begin{equation}
    \begin{aligned}
        \boldsymbol{t}^{(l+1)} &= \boldsymbol{t}_u^{(l+1)} \cap \boldsymbol{t}_i^{(l+1)}, \\
         Off(\boldsymbol{t}^{(l+1)})&=Min(\{Off(\boldsymbol{v}_1^{(l)}),\dots,Off(\boldsymbol{v}_{|\mathcal{N}_t|}^{(l)})\}),
    \end{aligned}
\end{equation}
where $\mathcal{N}_t$ is the neighbors of the tag $t$, $\boldsymbol{v}_k$ denotes the user or item box representations. Notably, we use intersection to integrate the $\boldsymbol{t}_u^{(l+1)}$ and $\boldsymbol{t}_i^{(l+1)}$ for extracting the shared traits between users and items, which can be recognized as $\boldsymbol{t}^{(l+1)}$.

\textbf{Item-aware Aggregation.}
In CTG, there are also two types of neighbors for item, i.e., tags and users. Tags can also serve as characteristics of item, with which we can form a specific item. Consistent with user aggregation, we utilize the union operation to aggregate the neighboring tag nodes:
\begin{equation}
    \begin{aligned}
        \boldsymbol{i}^{(l+1)}_t = \bigcup_{k=1}^ {|\mathcal{N}_i^t|}\boldsymbol{t}_k^{(l)}, \\
    \end{aligned}
\end{equation}
where $i^{(l+1)}_t$ is the item box representation from tag neighbors and $\mathcal{N}_i^t$ is the tag neighbor set of item $i$. 

When it comes to users, we still adopt union operation for two reasons: (1) To capture the multi-facet feature of the product. Different people may interact with the same item driven by various intents. By applying the union operation to user boxes, this diversity can be effectively represented. (2) To express the popularity of item. The more users who have purchased the item, the larger the volume of the item box becomes. Along this line, for any new-coming user, its box is more likely to be encompassed by the box of popular item, which indicates a higher probability of purchasing the item. 
\begin{equation}
    \begin{aligned}
        \boldsymbol{i}^{(l+1)}_u = \bigcup_{k=1}^ {|\mathcal{N}_i^u|}\boldsymbol{u}_k^{(l)}, \\
    \end{aligned}
\end{equation}
where $i^{(l+1)}_u$ is the item box representation from user neighbors and $\mathcal{N}_i^u$ is the user neighbor set of item $i$. 
In summary, the aggregation formula for the item is as follows:

\begin{equation}
    \begin{aligned}
        \boldsymbol{i}^{(l+1)} &= \boldsymbol{i}_t^{(l+1)} \cup \boldsymbol{i}_u^{(l+1)},, \\
         Off(\boldsymbol{i}^{(l+1)})&=Max(\{Off(\boldsymbol{v}_1^{(l)}),\dots,Off(\boldsymbol{v}_{|\mathcal{N}_i|}^{(l)})\}).
    \end{aligned}
\end{equation}
where $\mathcal{N}_i$ is the neighbors of the item $i$. After stacking $L$ layers, we will obtain the ultimate box representation $\boldsymbol{u}^{(L)}$ and $\boldsymbol{i}^{(L)}$, which are feed into next module to get the predictive result.


\subsection{Gumbel-based Volume Objective}
After obtaining the higher-order box representations, we emphasize that the key point turns to the similarity calculation between user and item. Obviously, it is intractable to still apply dot product to measure their similarity due to the complex structure within the box representations. Instead, we need to consider more sophisticated metrics that can capture the geometric relationships between these high-dimensional boxes, which aligns with the box properties. 

To this end, we utilize the volume of the intersection between user and item boxes as their similarity. However, we may face gradient vanishing issue if there is no intersection between two boxes. Inspired by \cite{DBLP:conf/nips/DasguptaBZV0M20}, we attempt to regard the acquired high-order box representation $\boldsymbol{v}^{(L)}$ as gumbel box, where minimum corner $\boldsymbol{z_{\boldsymbol{v}^{(L)}}}$ and maximum corner $\boldsymbol{Z}_{\boldsymbol{v}^{(L)}}$ follows Gumbel distribution: 
\begin{equation}
    \begin{aligned}
        f(x;\mu,\beta)=\frac{1}{\beta} exp(-\frac{x-\mu}{\beta} - e^{-\frac{x-\mu}{\beta}}),        
    \end{aligned}
\end{equation}
where $\beta$ controls the scale of the distribution, and $\mu$ governs the mean of the distribution. Note it is min/max stable, i.e., the min/max of two such variables still follows Gumbel distribution \cite{DBLP:conf/uai/BoratkoBDM21}. For simple notation, we omit the $L$ superscript in this subsection. Formally, we define the $\boldsymbol{z}_v$ and $\boldsymbol{Z}_v$ as: 

\begin{equation}
    \begin{aligned}
        \label{gumbel_distribution}
        \boldsymbol{z}_{v}  \sim Gumbel(\boldsymbol{\mu}^z_{v}, \beta), \boldsymbol{\mu}^z_{v} = Cen(\boldsymbol{v}) - Off(\boldsymbol{v}), \\
        \boldsymbol{Z}_{v}  \sim Gumbel(\boldsymbol{\mu}^Z_{v}, \beta), \boldsymbol{\mu}^Z_{v} = Cen(\boldsymbol{v}) + Off(\boldsymbol{v}).\\
    \end{aligned}
\end{equation}
According to min/max stability \cite{DBLP:conf/uai/BoratkoBDM21}, we then can derive the the minimum and maximum corners of the intersected box between user and item box:
\begin{equation}
    \begin{aligned}
        \boldsymbol{z}_{ui} & = Max(\boldsymbol{z}_{u},  \boldsymbol{z}_{i}) \sim Gumbel(\boldsymbol{\mu}^z_{ui}, \beta), \\ 
        \boldsymbol{Z}_{ui} & = Min(\boldsymbol{Z}_{u}, \boldsymbol{Z}_{i}) \sim Gumbel(\boldsymbol{\mu}^Z_{ui},\beta), \\
    \boldsymbol{\mu}^z_{ui} & = {\beta}ln(e^{\boldsymbol{\mu}_{u}^z/\beta}+e^{\boldsymbol{\mu}_{i}^z/\beta}),\\
    \boldsymbol{\mu}^Z_{ui} & = -{\beta}ln(e^{-\boldsymbol{\mu}_{u}^Z/\beta}+e^{-\boldsymbol{\mu}_{i}^Z/\beta}),
    \end{aligned}
\end{equation}
where $\boldsymbol{u}$ and $\boldsymbol{i}$ are the user and item box representation after stacking $L$ layers, respectively. Therefore, $\boldsymbol{\mu}^z_u$ and $\boldsymbol{\mu}^Z_u$, $\boldsymbol{\mu}^z_i$ and $\boldsymbol{\mu}^Z_i$ are separately the two corner embeddings of user and item box. Here, the derivation of $\boldsymbol{\mu}^z_{ui}$ and $\boldsymbol{\mu}^Z_{ui}$ can be found in \cite{DBLP:conf/nips/DasguptaBZV0M20}.
Next, in order to get the expected volume, we derived the formula for calculating the expected length for each dimension. Thus we arrive at the expected volume formulation as following: 
\begin{equation}
    \begin{aligned}
        \label{expected_volume}    
     \mathbb{E}[max(\boldsymbol{Z}_{ui}-\boldsymbol{z}_{ui},0)] = \prod^d_k{2{\beta}K_{0}\left(2e^{-({\mu^Z_{ui,k}-\mu^z_{ui,k}}) /2\beta}\right)},
    \end{aligned}
\end{equation}
where $d$ is the dimension of embedding, $\mu_{ui,k}^z$ is the $k$-th element in $\boldsymbol{\mu}_{ui}^z$, and $\mu_{ui,k}^Z$ follows the same principal. $K_0$ is the modified Bessel function of second kind, order 0. The proof of this statement is given in \cite{DBLP:conf/nips/DasguptaBZV0M20}. 

To further analyse the Equation (\ref{expected_volume}), let $m(x)=2\beta{K_0}\left(2e^{x/2\beta}\right)$, then the $m(x)$ is essentially exponential as $x$ increases, and the volume function approaches a hinge function as $\beta \rightarrow 0$,  which leads to numerical stability concerns. Here, we use softplus-like function to replace the $m(x)$
\begin{equation}
    \begin{aligned}
    m(x)={\beta}log\left(1+e^{x/\beta - 2\gamma}\right),
    \end{aligned}
\end{equation}
where $\gamma$ is Euler-Mascheroni constant. Equipped with this, we have our final version of volume forumulation:
\begin{equation}
    \begin{aligned}
    \label{gumbel_volume}
    Vol(Box_{ui}) &= \prod^d_k{{\beta}log\left(1+e^{-(\mu_{ui,k}^Z - \mu_{ui,k}^z)/\beta - 2\gamma}\right)}.
    \end{aligned}
\end{equation}

\subsection{Model Training}
Through the above modules, the preference score for a user $u$ toward an item $i$ is defined as the volume of intersection of two box representation:
\begin{align}
    \hat{y}_{ui} &= log(Vol(Box_{ui})).
\end{align}
Here, we use a log function to prevent gradient vanishing. Then we employ Bayesian Personalized Ranking (BPR) loss to train the whole model, which assumes that observed interactions should receive higher scores than unobserved one:
\begin{align}
    \label{eq:final_loss}
    \mathcal{L} = \sum_{(u, i^+, i^-) \in E} -\log \sigma (\hat{y}_{ui^+} - \hat{y}_{ui^-}) + \lambda \Vert \Theta \Vert^2,
\end{align}
where $E$ is the set of training interactions, $i^+$ and $i^-$ are separately positive and negative samples from training data,  $\lambda \Vert \Theta \Vert ^2$ is the regularization term to prevent overfitting.


\subsection{Model Analysis}
We conducted a comparative analysis with existing models to demonstrate the rationale of BoxGNN.
\subsubsection{Relation with GAT} GAT \cite{DBLP:conf/iclr/VelickovicCCRLB18} is a well-known scheme that utilizes an attention mechanism to gauge the significance of neighboring signals to recognize more vital information. Formally, The graph aggregation formulation in GAT can be defined as:
\begin{align} 
    \boldsymbol{h}^{(l)}_i  = \sigma \left( \sum_{j\in {\mathcal{N}_i}}\alpha_{ij} \boldsymbol{W} \boldsymbol{h}^{(l-1)}_j \right),
    \alpha_{ij} = \frac{exp(\boldsymbol{h}_j^{(l-1)})}{\sum_{k \in \mathcal{N}_i} exp(\boldsymbol{h}_k^{(l-1)})},
\end{align}
where $\boldsymbol{W}$ is the learnable parameter and $\sigma$ is the activation function. Comparing with Equation \ref{equation_2}, if we set the offset of our boxes to zero, then our aggregation operation would closely mirrors GAT.
\begin{table}[t]
\caption{Statistics of three real-world datasets.}
\label{tab:statistics}
\begin{tabular}{lcccc}
\hline
Dataset & \#Users & \#Items & \#Tags & \#Assignment  \\
\hline
MovieLens & 1,651      &  5,381 & 1,586       & 36,728        \\
LastFm&  1,808  & 12,212     & 2,305       &   175,641      \\
E-shop & 7,277 & 26,726 & 4,146       & 237,059        \\

\hline
\end{tabular}
\end{table}
\begin{table*}[t]
\caption{The experimental comparison among a wide range of recommendation approaches for three datasets. R@K and N@K stand for Recall@K and NDCG@K, respectively. BoxGNN w/o tags represents the results of our BoxGNN running on a dataset from which the tags information has been removed. The best results are in bold and the secondary best results are underlined. * indicates the statistically significance over the best baseline using t-test with $p< 0.05$ .}
\label{tab:overall_performance}
\setlength{\tabcolsep}{1.5mm}{
\begin{tabular}{p{2.3cm}|cccc|cccc|cccc}
\hline

\multirow{2}{*}{Model}                  & \multicolumn{4}{c|}{MovieLens}              &  \multicolumn{4}{c|}{LastFm}                 &  \multicolumn{4}{c}{E-shop} 
\\
& R@10& R@20 & N@10&N@20 &R@10& R@20 & N@10&N@20&R@10& R@20 & N@10&N@20    \\
\hline
BPR & 0.0453 & 0.0661 & 0.0320 & 0.0380 & 0.0673 & 0.0978 & 0.0563 & 0.0643 & 0.3409 & 0.4491 & 0.2632 & 0.3006
\\
NFM & 0.0351 & 0.0608 & 0.0225 & 0.0306 & 0.0762 & 0.1173 & 0.0656 & 0.0768 & 0.0476 & 0.1230 & 0.0285 & 0.0536
\\
IFM & 0.0338 & 0.0474 & 0.0191 & 0.0232 & 0.0562 & 0.0898 &  0.0412 & 0.0507 & 0.1413 & 0.2279 & 0.0961 & 0.1255
\\
\hline
LightGCN & 0.0446 & 0.0724 & 0.0313 & 0.0393 & 0.1020 & 0.1555 & 0.0875 & 0.1002 & 0.2749 & 0.3710 & 0.2118 & 0.2446
\\
NGCF & 0.0352 & 0.0592 & 0.0273 & 0.0331 & 0.1149 & 0.1580 &  0.0943 & 0.1041 & \underline{0.3748} & \underline{0.4976} & \underline{0.2852} & \underline{0.3283} 
\\
KGIN & 0.0633 & 0.0978 & 0.0516 & 0.0607 & 0.1113 & 0.1662 & 0.0924 & 0.1068 & 0.2934 & 0.3900 & 0.2300 & 0.2619
\\
KGRec & 0.0468 & 0.0819 &0.0380  & 0.0476 & 0.1166 & 0.1665 & 0.0996 & 0.1116 & 0.2681 & 0.3515 & 0.2118 & 0.2394
\\
\hline
DSPR & 0.0661 & 0.0929 & 0.0476 & 0.0549 & 0.0505 & 0.0885  & 0.0438 & 0.0541 & 0.2336 & 0.3297 & 0.1686 & 0.2021
\\
LFGCF & \underline{0.0812}&	\underline{0.1167}&	\underline{0.0601}&	\underline{0.0705}&	\underline{0.1241}&	\underline{0.1848}&	\underline{0.1058}&	\underline{0.1198}&	0.3586&	0.4803&	0.2598&	0.3018 
\\
\hline
BoxGNN w/o tags& 0.0533 &	0.0794&0.0362&0.0445&0.1271&	0.1814&0.1023&0.1156   & 0.4260  & 0.5432 & 0.3298  & 0.3718 
\\
BoxGNN & \textbf{0.0866} &	\textbf{0.1226}&\textbf{0.0704}*&\textbf{0.0812}*&\textbf{0.1350}*&	\textbf{0.1934}*&\textbf{0.1124}&\textbf{0.1264}   & \textbf{0.4505}*  & \textbf{0.5744}* & \textbf{0.3447}*  & \textbf{0.3881}* 
\\
\hline
\end{tabular}
}

\end{table*}
\section{EXPERIMENTS}
\subsection{Experimental Settings}
\subsubsection{Data Description.} To demonstrate the superiority of proposed BoxGNN, we conduct experiments on two publicly available benchmarking datasets and one LLM-enhanced e-commercial dataset, i.e. MovieLens, LastFm and E-shop. Notably, Movielens and LastFm are released in HetRec 2011\footnote{https://grouplens.org/datasets/hetrec-2011.} and E-shop would be publicly available after internal review. 
\begin{itemize}
    \item \textbf{MovieLens:} This is a movie recommendation collection released by the GroupLens research group\footnote{http://www.grouplens.org.}. Within this dataset, each user is associated with a list of tag assignments corresponding to the movies they have interacted with.
    \item \textbf{LastFm:} This is an artist recommendation dataset obtained from the online music system Last.FM\footnote{http://www.last.fm.com}. In this dataset, each user has a list of tags assigned to artists.
    \item \textbf{E-shop:} This is the real-world commercial dataset collected from online scenario. By feeding the purchased item titles into Large Language Models(LLMs), such GPT-4 turbo, we ask to identify the underlying interests that led to the purchase behaviour. Then LLMs generate a list of tags which are assigned with the users and items.
    
\end{itemize}

Following existing literature\cite{DBLP:conf/cikm/ChenGTXDHW20}, to ensure the data quality, we filter out the infrequent tags that are used less than 5 times in three dataset. The statistics of datasets are summarized in Table \ref{tab:statistics}. We randomly select 80\%, 10\%,10\% of the assignments as training set, validation set and test set, respectively. It is worth noting that we construct the CTG by the users, items and tags in training data.

\subsubsection{Evaluation Metrics.} We utilize two representative evaluation metrics for top-K recommendation, i.e. $Recall@K$ and $NDCG@K$ (Normalized Discounted Cumulative Gain) across all experiments, where $K=10,20$. We adopt all-ranking strategy in validation and test set. We evaluate on the validation set every 5 epochs, and we perform early stopping if there is no improvement for 10 consecutive evaluations. The performance of the best model on the test set is considered our final result.

\subsubsection{Compared baselines.} 
To demonstrate the effectiveness of proposed BoxGNN, we gather recommendation techniques from various domains as baselines, involving classic method BPR\cite{DBLP:conf/uai/RendleFGS09}, feature-based (NFM\cite{DBLP:conf/sigir/0001C17},IFM\cite{DBLP:conf/ijcai/YuWY19}), GNN-based (LightGCN\cite{DBLP:conf/sigir/0001DWLZ020}, NGCF\cite{DBLP:conf/sigir/Wang0WFC19}), KG-based (KGIN\cite{DBLP:conf/www/WangHWYL0C21}, KGRec\cite{DBLP:conf/kdd/YangHXH23}) and tag-based (DSPR\cite{DBLP:conf/ijcai/XuLCMM17}, LFGCF\cite{DBLP:journals/corr/abs-2208-03454}) models.

\subsection{Experimental Results}
\subsubsection{Overall Performance} In this part, we conduct experiments on three datasets to show the superiority of our proposed BoxGNN compared with various baselines. From the results listed in Table. \ref{tab:overall_performance}, we have following observations: \\
\begin{itemize}
    \item Our proposed BoxGNN outperforms all of baselines across all datasets. This performance boost can be attributed primarily to two factors: On one hand, by modeling nodes as box embeddings, we can capture user uncertainty and diversity, which in turn enhances the performance of the recommendation system. On the other hand, by simulating a message propagation mechanism, nodes embeddings are enriched as they aggregate more high-order information.
    \item In most cases, GNN-based methods surpass feature-based methods, as the latter do not take advantage of their strengths when only tags are used as the feature. On the other hand, GNN-based approaches can effectively reap benefits from the graph structure and collaborative signals, resulting in more expressive representations.
    \item KG-based methods perform better on the Movielens and Lastfm datasets compared to GNN-based methods, because KG-based approaches incorporate tag information, which can assist the model in better capturing item characteristics and user interests. 
    \item BPR and NGCF demonstrate remarkable performance on the E-shop dataset compared to other baselines, which actually suggests that the user-item collaborative signal is adequate for user modeling. However, the additional introduction of tag information has weakened the performance of many models, indicating that there exists noise in the tag set within the E-shop dataset. This is understandable given that these tags are produced by LLMs.
    \item The performance of BoxGNN significantly surpasses that of BoxGNN w/o tags across three datasets, highlighting the critical importance of tag information in user-item matching task. Moreover, even after the removal of tags, BoxGNN retains its superior performance over both LightGCN and NGCF, which demonstrates the capability of our approach in uncovering user interests.
\end{itemize}

\subsubsection{Ablation Study}
In this part, we conduct the ablation study to demonstrate the effect of each module on the overall model and deliver some insight on how they affect the results. 

\textbf{Effect of GNNs.} The GNN module is the cornerstone of our approach, which captures rich semantic signals from high-order neighbors. To show the effectiveness of this module, we create the model by setting $l=0$, which means the removal of GNN module, denoted as w/o GNN. From Table \ref{tab:ablation}, it is evident that there is a dramatic decrease in performance after removing the GNN module. This is because the box representation fail to absorb knowledge from high-order neighbors, resulting in poorer representation performance. This reaffirms the importance of high-order signals and the effectiveness of the GNN module in capturing these signals.

\textbf{Effect of Gumbel-based Volume Objective.} This objective is the key component to train the whole process, which has the ability to consistently deliver gradient signals. To demonstrate its superiority, we obtain a variant that adopts the direct calculation of the volume for the intersected box, namely, w/o Gumbel. As shown in Table \ref{tab:ablation}, After removing the Gumbel-based module, our performance suffered a significant decline. The reason lies in the alleviation of the gradient vanishing problem when there is no overlap between two boxes.
\begin{table}[t]
\caption{Ablation studies for investigating the effects of each module.}
\label{tab:ablation}
\setlength{\tabcolsep}{1mm}{
\begin{tabular}{c|cc|cc|cc}
\hline
                                     & \multicolumn{2}{c|}{MovieLens}        & \multicolumn{2}{c|}{LastFm}        & \multicolumn{2}{c}{E-shop}     \\
                                         & Recall & NDCG & Recall & NDCG & Recall & NDCG \\ \hline
 w/o GNN &  0.0343&	0.0237&	0.0965&	0.0840&	0.3843&	0.2961     \\ 
                          w/o Gumbel     &  0.0652&	0.0544&	0.1083&	0.0905&	0.4040&	0.3166    \\
                          all            &   \textbf{0.0866}     & \textbf{0.0704}     &  \textbf{0.1350}      &  \textbf{0.1124}    &  \textbf{0.4490}      &  \textbf{0.3455}    
\\ \hline
\end{tabular}
}

\end{table}

\begin{figure*}[htbp]
\centering
\caption{Effect of the parameter $\beta$ on all datasets.}
\vspace{1.8mm}
\label{fig.params.beta}
\includegraphics[width=2.0\columnwidth]{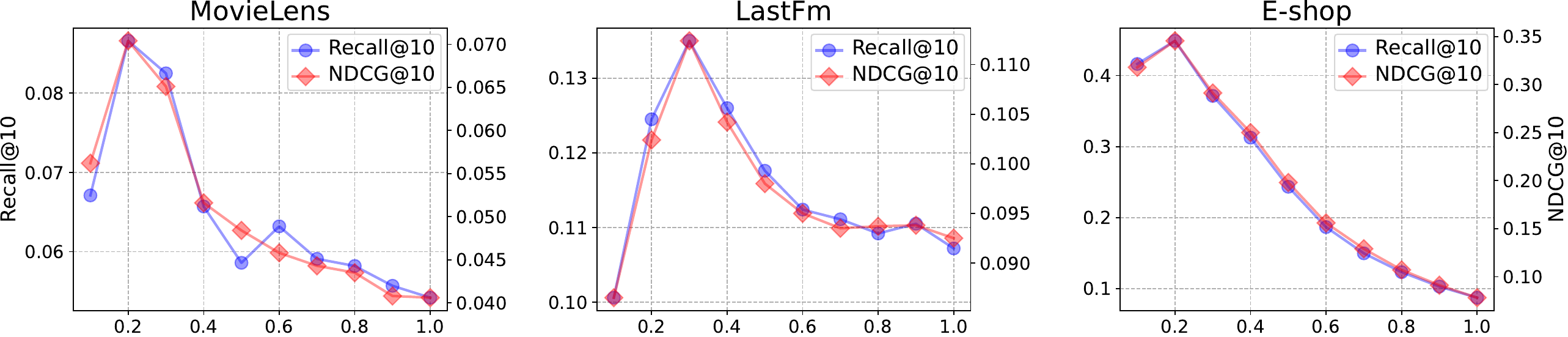} 

\label{paramters}
\end{figure*}



\begin{table}[t]
\caption{Impact of the number of layers.}
\label{tab:layers}
\setlength{\tabcolsep}{1mm}{
\begin{tabular}{c|cc|cc|cc}
\hline
                                     & \multicolumn{2}{c|}{MovieLens}        & \multicolumn{2}{c|}{LastFm}        & \multicolumn{2}{c}{E-shop}     \\
                                         & Recall & NDCG & Recall & NDCG & Recall & NDCG \\ \hline
 BoxGNN-1 &  0.0474&	0.0355&	0.1175&	0.0964&	0.4203&	0.3251     \\ 
                          BoxGNN-2     &  0.0681&	0.0563&	0.1254&	0.1050&	\textbf{0.4490}&	\textbf{0.3455}    \\
                          BoxGNN-3            &   \textbf{0.0866}     & \textbf{0.0704}     &  \textbf{0.1350}      &  \textbf{0.1124}    &  0.4343&	0.3301    
\\ \hline
\end{tabular}
}

\end{table}

\subsubsection{Impact of Model Depth} Furthermore, we turn to the influence of the layer number on the overall performance. The layer number is selected in the range of \{1, 2, 3\}, which is consistent with most GNN-based methods. As depicted in Table \ref{tab:layers}, we draw following observations: (1) The performance of BoxGNN-2 always exceeds that of BoxGNN-1, indicating that modeling high-order connectivity has a significant positive impact on the overall model performance. (2) Majority of datasets achieves the highest performance in BoxGNN-3, e.g., MovieLens and LastFm, it means that representation of absorbed third-order signals is more expressive than second-order representations. (3) Under the E-shop dataset, the performance of BoxGNN-3 is worse than that of BoxGNN-2, which is contrary to the previous two datasets. The reason is that the tags contain noise in this dataset, and excessive connectivity can introduce too much noise, thus resulting in performance loss.
\subsubsection{Parameters Sensitivity} Moreover, we investigate the effect of the $\beta$ on the whole system. Note $\beta$ controls the scale of the Gumbel distribution. 
Similar to the temperature coefficient in the contrastive loss, when $\beta$ is larger, the Gumbel distribution is closer to a uniform distribution. In contrast, when the $\beta$ is smaller, its probability density function approaches the hinge function, meaning the random variable will degenerate into a constant. In this paper, we need to strike a balance between the distinctiveness among the boxes and the uncertainty within each box. 

From Fig \ref{fig.params.beta}, we have following findings: (1) All the value of $\beta$ are around 0.2. In this line, we can ensure uncertainty within each box while still maintaining distinctiveness among the boxes. This is consistent with the principle mentioned above. (2) With the increase in $\beta$, performance of our BoxGNN rapidly declines. This is reasonable that as $\beta$ increases, the Gumbel distribution tends towards a uniform distribution, resulting in the gradual disappearance of differences among boxes, making it impossible to accurately model the users. (3) As $\beta$ decreases,  performance of our system will also be affected to some extent. This is inevitable because as $\beta$ decreases, the uncertainty of the Gumbel distribution gradually disappears, leading to the reemergence of the vanishing gradient problem, thereby resulting in a drop in model accuracy.
\subsubsection{Visualization}
\begin{figure}[htbp]
\centering
\caption{The visualization of centroid points of Box and node embeddings of LFGCF on MovieLens dataset.}
\label{fig.visualization}
\vspace{3mm}
\begin{subfigure}{0.22\textwidth}
    \centering
    \includegraphics[width=1\textwidth]{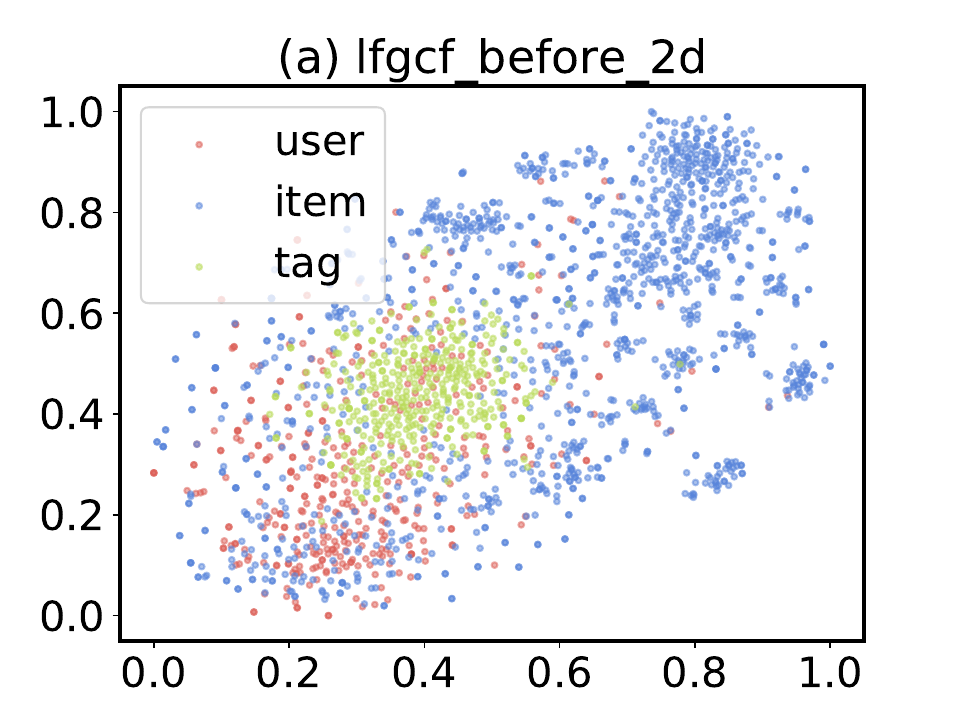}
\end{subfigure}
\begin{subfigure}{0.22\textwidth}
    \centering
    \includegraphics[width=1\textwidth]{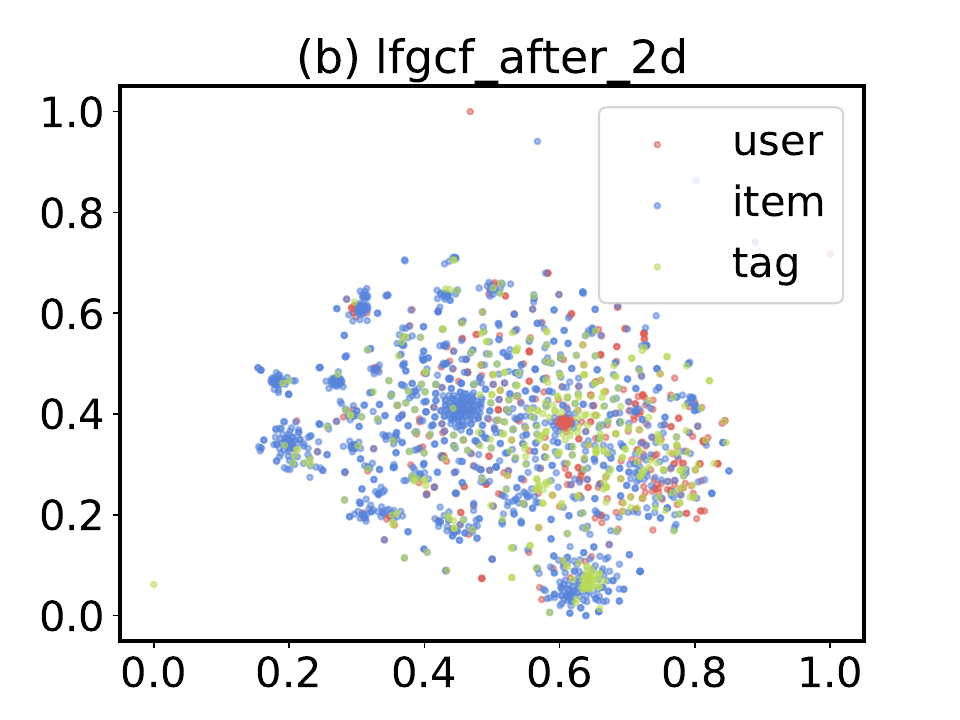}
\end{subfigure}

\begin{subfigure}{0.22\textwidth}
    \centering
    \includegraphics[width=1\textwidth]{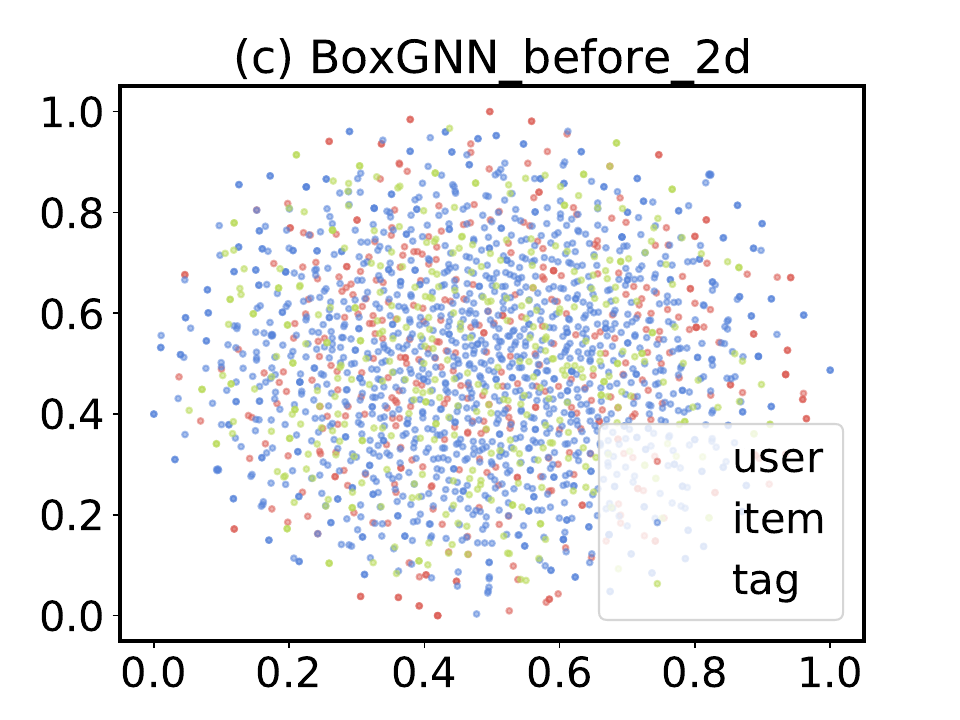}
\end{subfigure}
\begin{subfigure}{0.22\textwidth}
    \centering
    \includegraphics[width=1\textwidth]{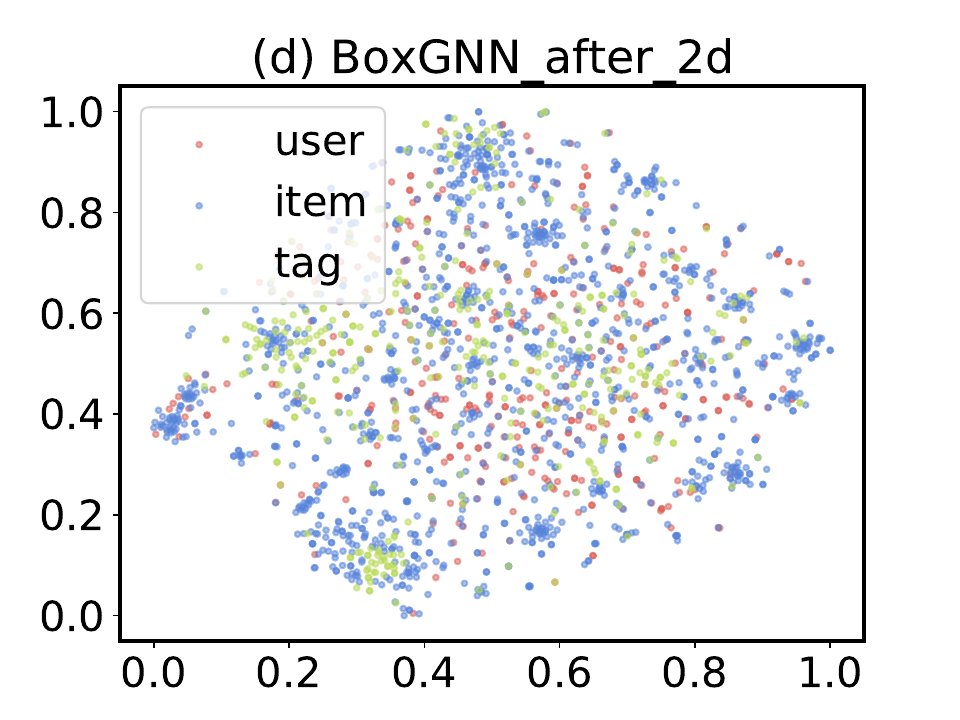}
\end{subfigure}
\end{figure}
We visualized the distribution of the centroid points of Box on the Movielens dataset. For comparison, we also visualized the distribution of node embeddings from LFGCF. From the Figure \ref{fig.visualization} (a), we can see that before performing GNN, nodes of the same type in the LFGCF model are close to each other. This indicates that they have a high similarity among nodes of the same type, lacking differentiation and diversity. Furthermore, from the Figure \ref{fig.visualization} (b), it is apparent that after performing GNN, the clustering phenomenon becomes more significant.
From the Figure \ref{fig.visualization} (c), we can observe that before performing GNN operations, the distribution of points in BoxGNN is quite uniform. This indicates that our method possess sufficient diversity and differentiation, and this part of the gain mainly comes from Box modeling, which aligns with our motivation for using Box to model diversity of user interests. After the GNN operation, Figure \ref{fig.visualization} (d) shows several blue clustering points (items), which is quite reasonable since the purpose of GNN itself is to explicitly aggregate nodes with similar information. Furthermore, the distribution of these clustering points is still quit uniform, rather than forming a few large clustering points like in Figure \ref{fig.visualization} (b).


\section{Conclusion}
In this paper, we proposed a novel Box-based graph neural network for tag-aware recommendation which models the uncertainty and diversity of user interests and thus boosts the overall performance. Specifically, we first formulate the box embedding in the vector space and initialize all nodes as box embeddings. Next, with two logical operations, we perform the type-aware message aggregation to obtain the high-order box representations. Then, to avoid gradient vanishing issue, we devise the gumbel-based volume objective to refine the representation of boxes. Finally, extensive experiments on three real-world datasets demonstrated the superiority of the BoxGNN compared with competitive baselines. 
\begin{acks}
This work was supported in part by the grants from National Natural Science Foundation of China (No.62222213, 62202443, 62207031, 62072423).
\end{acks}

\bibliographystyle{ACM-Reference-Format}
\bibliography{sample-base}

\appendix
\section{Appendix}

\subsection{Baseline Details}
\begin{itemize}
    \item \textbf{BPR\cite{DBLP:conf/uai/RendleFGS09}:} It is a well-known  model that employs Bayesian personalized ranking as the objective function.
    \item \textbf{NFM\cite{DBLP:conf/sigir/0001C17}:} It is the first work to model the interactions among high-order features via neural network.
    \item \textbf{IFM\cite{DBLP:conf/ijcai/YuWY19}:} It is a variant of factorization matrix method to consider the impact of each individual input upon the representation of features.
    \item \textbf{NGCF\cite{DBLP:conf/sigir/Wang0WFC19}:} It first propagate embeddings of user and items in an explicit manner to capture the expressive high-order collaborative signals.
    \item \textbf{LightGCN\cite{DBLP:conf/sigir/0001DWLZ020}:} It is the state-of-the-art GNN-based approach with the simplest framework that only preserves the neighborhood aggregation part.
    \item \textbf{KGIN\cite{DBLP:conf/www/WangHWYL0C21}:} It considers the user-item interaction at the finer granularity of intents and utilizes the relational path-aware aggregation mechanism to identify user intents. 
    \item \textbf{KGRec\cite{DBLP:conf/kdd/YangHXH23}:} It is the self-supervised KG-based scheme that applies mask mechanism to locate the critical information and discard the noisy and irrelevant nodes in KG.
    \item \textbf{DSPR\cite{DBLP:conf/ijcai/XuLCMM17}:} It is the first model to employs MLPs to process tag-based features to extract user and item representations.
    \item \textbf{LFGCF\cite{DBLP:journals/corr/abs-2208-03454}:} It is the latest state-of-the-art method in TRS, which borrows the idea of LightGCN and separately models three types of message aggregation.
\end{itemize}
\subsection{Reproductive Settings}
We implement our BoxGNN in PyTorch. For a fair comparison, we fix the embedding size to 64 for all methods. The batch size is fixed to 1024. We employ the grid search strategy to optimize the hyper-parameters with Adam optimizer. In detail, the learning rate is searched amongst $\{1e^{-5}, 1e^{-4}, 1e^{-3}\}$, the regularization coefficient is tuned in $\{1e^{-6}, 1e^{-5}, 1e^{-4}, 1e^{-3}\}$. We leverage node dropout to avoid over-fitting, where dropout ratio is tuned amongst $\{0.1, 0.2, ..., 0.7\}$. Notably, we use xavier\cite{DBLP:journals/jmlr/GlorotB10} to initialize both center and offset embeddings.
\end{document}